\documentclass[lettersize,journal]{IEEEtran}
\usepackage{amsmath,amsfonts, amssymb}
\usepackage{algorithmic}
\usepackage{algorithm}
\usepackage{array}
\usepackage[caption=false,font=normalsize,labelfont=sf,textfont=sf]{subfig}
\usepackage{textcomp}
\usepackage{stfloats}
\usepackage{url}
\usepackage{verbatim}
\usepackage{graphicx}
\usepackage{cite}
\usepackage{orcidlink}

\newcommand{\uhfone}{220}
\newcommand{\uhftwo}{280}
\newcommand{\nummods}{25}

\begin{document}

\title{The Simons Observatory: Characterization of the 220/280 GHz TES Detector Modules}

\author{Daniel Dutcher\orcidlink{0000-0002-9962-2058},
Peter Dow\orcidlink{0009-0006-8427-6259},
Shannon M. Duff\orcidlink{0000-0002-9693-4478},
Shawn W. Henderson\orcidlink{0000-0001-7878-4229},
Johannes Hubmayr\orcidlink{0000-0002-2781-9302},
Bradley R. Johnson\orcidlink{0000-0002-6898-8938},
Michael J. Link\orcidlink{0000-0003-2381-1378},
Tammy J. Lucas\orcidlink{0000-0001-7694-1999},
Michael D. Niemack\orcidlink{0000-0001-7125-3580},
Yudai Seino\orcidlink{0000-0001-5680-4989},
Rita F. Sonka\orcidlink{0000-0002-1187-9781},
Suzanne Staggs\orcidlink{0000-0002-7020-7301},
Yuhan Wang\orcidlink{0000-0002-8710-0914},
Kaiwen Zheng\orcidlink{0000-0003-4645-7084}
% ~\IEEEmembership{Staff,~IEEE,}
        % <-this % stops a space
\thanks{D. Dutcher, Y. Seino, R. F. Sonka, S. Staggs, Y. Wang, and K. Zheng conducted this work with the Department of Physics, Princeton University, Princeton, NJ 08544 USA (email: ddutcher@princeton.edu).
R. F. Sonka is now with the Astrophysics Science Division, NASA Goddard Space Flight Center, Greenbelt, MD 20771 USA.
Y. Wang is now with the Department of Physics, Cornell University, Ithaca, NY, 14853, USA.
P. Dow and B. R. Johnson are with the Department of Astronomy, University of Virginia, Charlottesville, VA 22904 USA.
S. M. Duff, J. Hubmayr, M. J. Link and T. J. Lucas are with the Quantum Sensors Division, National Institute of Standards and Technology, Boulder, CO 80305 USA.
S. W. Henderson is with the Kavli Institute for Particle Astrophysics and Cosmology, Stanford, CA 94305 USA and also with SLAC National Accelerator Laboratory, Menlo Park, CA 94025 USA.
M. D. Niemack is with the Department of Physics and also with the Department of Astronomy, Cornell University, Ithaca, NY, 14853, USA.
}% <-this % stops a space
}

\maketitle

\begin{abstract}
The Simons Observatory (SO) is a new suite of cosmic microwave background telescopes in the Chilean Atacama Desert with an extensive science program spanning cosmology, Galactic and extragalactic astrophysics, and particle physics.
SO will survey the millimeter-wave sky over a wide range of angular scales using six spectral bands across three types of dichroic, polarization-sensitive transition-edge sensor (TES) detector modules: Low-Frequency (LF) modules with bandpasses centered near 30 and 40~GHz, Mid-Frequency (MF) modules near 90 and 150~GHz, and Ultra-High-Frequency (UHF) modules near 220 and 280~GHz.
Twenty-five UHF detector modules, each containing 1720 optically-coupled TESs connected to microwave SQUID multiplexing readout, have now been produced.
This work summarizes the pre-deployment characterization of these detector modules in laboratory cryostats.
Across all UHF modules, we find an average operable TES yield of 83\%, equating to over 36,000 devices tested.
The distributions of (220, 280)~GHz saturation powers have medians of (24, 26)~pW, near the centers of their target ranges.
For both bands, the median optical efficiency is 0.6, the median effective time constant is 0.4~ms, and the median dark noise-equivalent power (NEP) is $\sim$40~aW$/\sqrt{\mathbf{Hz}}$.
The expected photon NEPs at (220, 280)~GHz are (64, 99)~aW$/\sqrt{\mathbf{Hz}}$, indicating these detectors will achieve background-limited performance on the sky.
Thirty-nine UHF and MF detector modules are currently operating in fielded SO instruments, which are transitioning from the commissioning stage to full science observations.
\end{abstract}


\section{Introduction}
%\IEEEPARstart{R}{ecent} results from ground-based observations of the cosmic microwave background (CMB), such as those from BICEP, SPT, ACT, and CLASS, have continued to make developments at the forefront of cosmology, addressing the science of inflation, reionization, structure formation, and galactic and extragalactic astrophysics.

\IEEEPARstart{T}{he} Simons Observatory (SO) is a new cosmic microwave background project sited in the Atacama Desert at an elevation of 5200~m, with a current instrument suite of three 0.42~m-diameter small-aperture telescopes (SATs) and one 6~m-primary large-aperture telescope (LAT).
The science program of SO is broad, spanning cosmology, Galactic and extragalactic astrophysics, and particle physics.
The LAT will conduct a wide-field survey, mapping over 70\% of the sky with arcminute resolution, while the SATs will focus on a smaller sky area for a targeted search for the signature of cosmic inflation \cite{thesimonsobservatorycollaboration2019a, thesimonsobservatorycollaboration2025}.

SO will observe across six spectral bands centered approximately at (30, 40, 90, 150, 220, 280)~GHz.
This spectral coverage is achieved by using three types of dichroic, polarization-sensitive transition-edge sensor (TES) detector arrays: Low-Frequency (LF) arrays at 30/40~GHz, Mid-Frequency (MF) at 90/150~GHz, and Ultra-High-Frequency (UHF)\footnotemark{} arrays at \uhfone{}/\uhftwo{}~GHz.
The detector arrays are combined with optical coupling elements and multiplexing circuitry to form integrated detector modules.
Each SAT is populated with seven detector modules, while the LAT can accommodate up to 39 modules.% distributed across 13 modular optics tubes.

\footnotetext{
``UHF" is used as opposed to ``HF", as the latter is used by the ACT collaboration to refer to a combination of 150/230~GHz bands \cite{henderson2016}. 
}

%, corresponding to two SATs and four LAT optics tubes,
The pre-deployment characterization of 26 MF detector modules is reported on in \cite{dutcher2024}, while the design and testing of the first prototype UHF detector module can be found in \cite{healy2022}. 
Here, we summarize the laboratory characterization of \nummods{} production-design UHF detector modules, from which 19 will be selected for installation into SO instruments.
Thirteen UHF modules have already been deployed to the field in one SAT and the LAT, and an additional six modules will soon be installed in the LAT as part of the Advanced SO upgrade \cite{thesimonsobservatorycollaboration2025}.

\section{Module Design}

\begin{figure*}[!t]
\centering
\includegraphics[width=6.5in]{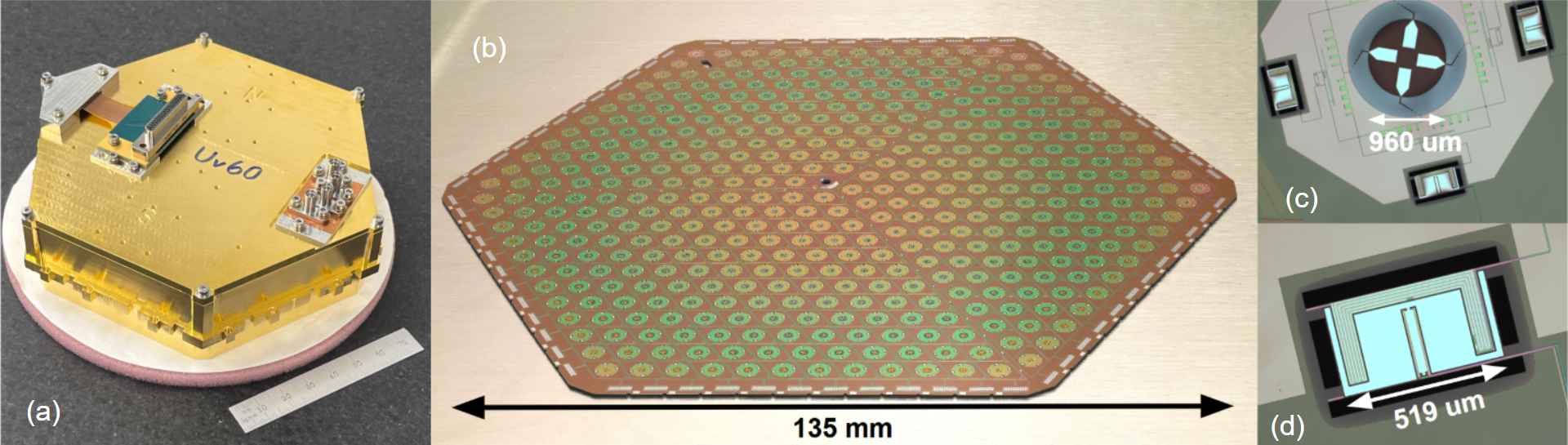}
\caption{
An assembled SO UHF detector module is shown in (a), containing a detector wafer, optical coupling components, and multiplexing circuitry. A bare detector wafer is shown in (b)¸ with insets depicting a single pixel (c) and one TES bolometer (d).
The bolometer consists of a suspended, thermally-isolated island onto which power from the antenna is deposited via a lossy meander. The rectangular structure running down the center of the island is the TES itself.
}
\label{fig:UHF_UFM_wafer}
\end{figure*}

The SO UHF and MF detector modules share a common design \cite{mccarrick2021a,healy2020}, containing 430 feedhorn-coupled orthomode transducers that each connect to four AlMn TES bolometers, corresponding to two orthogonal linear polarizations in two frequency bands.
Each module therefore has a total of 1720 antenna-coupled TESs, plus an additional 36 dark TESs used for calibration.
A UHF detector module and aspects of a detector wafer are depicted in Figure~\ref{fig:UHF_UFM_wafer}.
Details on NIST detector wafer fabrication can be found in \cite{duff2024}.
The TESs are read out through a microwave SQUID multiplexing ($\mu$mux) \cite{mates2011} architecture, in which each DC voltage-biased TES couples to an rf-SQUID, which in turn is coupled to a GHz resonator.
Probe tones interrogate the resonators, which shift with incident sky signal.
The SO $\mu$mux circuit is operated with SLAC Microresonator RF (SMuRF) warm readout electronics \cite{yu2023}.

The rf-SQUIDs and resonators are patterned from Nb on NIST multiplexer chips \cite{dober2021}, with each chip containing 65 SQUID-coupled channels and one ``dark" resonator channel.
To read out its 1756 TESs, each UHF and MF detector module uses two RF readout chains, each spanning 4--6~GHz and containing 14 daisy-chained multiplexer chips.
%, for a total of 924 resonators, 910 of which are coupled to SQUIDs, and $\sim880$ of which are connected to TESs.
To bias each TES to its operating point in the superconducting transition, $\sim$130--170 TESs are wired in series and supplied a common DC bias, with each detector module containing twelve such DC bias lines.
Each TES is wired in parallel with an $R_{sh} \approx 400~\mu\Omega$ shunt resistor.
A DC/RF routing wafer houses the shunt resistors and serves as the interface between the detector wafer, multiplexer chips, and external connectors.

\section{Testing Method}
The experimental setup and testing method are identical to those employed in \cite{dutcher2024} and described in \cite{wang2022a}.
Before coupling to a detector wafer, elements of the multiplexing circuit are screened at NIST \cite{jones2024, whipps2023} and Cornell, and full multiplexing modules are assembled and cryogenically tested at Princeton to evaluate their yield, resonator properties, and readout noise-equivalent current (NEI) distributions.
A multiplexing module is then coupled to a detector wafer and optical coupling components to form a detector module.
Up to three modules are installed in a dilution refrigerator cryostat along with an adjustable-temperature black body source, or ``cold load."
One third of the detectors on each module are exposed to the cold load, while the remaining two thirds are masked.
Masking a detector module serves multiple purposes: it permits measurement of both dark and optical properties of the detectors in a single cool down of the cryostat, limits non-uniform illumination from the cold load, and provides a means by which to separate the detectors' optical response to the cold load from their thermal response to a potentially varying detector wafer temperature.

The principle method for taking TES data is to adjust the input bias voltage while measuring the current through the TES as it moves through its superconducting transition, resulting in an \mbox{$I$-$V$} curve.
Sets of \mbox{$I$-$V$} curves are taken as the cryostat cold stage is ramped from 60~mK to 200~mK and as the cold load is ramped from 9~K to 20~K; these sets of data are then analyzed to produce measurements of TES superconducting critical temperature $T_c$, normal resistance $R_N$, saturation power $P_\mathrm{sat}$,\footnotemark{} and optical efficiency $\eta_\mathrm{opt}$.
With the TESs biased to fixed points in their transitions, square-wave modulations are input on the TES bias lines, and the detectors' responses are analyzed to derive their effective time constants $\tau_\mathrm{eff}$ and responsivities, the latter of which are used to calculate the detector noise-equivalent power (NEP).

\footnotetext{
As in \cite{dutcher2024}, we evaluate $P_\mathrm{sat}$ at $50\%R_N$, where $P_\mathrm{sat}$ is the electrical bias power required to bias a TES in its superconducting transition at a bath temperature of 100~mK and zero optical loading.
}

\section{Results}

The detector module properties are summarized in Table~\ref{tab:det_params} and discussed below, following a description of the resonator performance.

\begin{table}[!h]
\caption{TES Bolometer Properties \label{tab:det_params}}
\centering
\begin{tabular}{lcc}
\hline
Parameter & 220~GHz & 280~GHz \\
\hline
  Yield &
  $82\%\pm4\%$ &
  $83\%\pm3\%$
\\
  $R_N$ &
  $7.7\pm0.3~\mathrm{m}\Omega$ &
  $7.7\pm0.3~\mathrm{m}\Omega$
\\
  $T_c$ &
  $163\pm7$~mK &
  $163\pm8$~mK
\\
  $P_\mathrm{sat}$ &
  $24\pm4$~pW &
  $26\pm4$~pW
\\
  $\mathrm{NEP}_\mathrm{dark}$ &
  $39\pm4~\mathrm{aW}/\sqrt{\mathrm{Hz}}$ & 
  $42\pm7~\mathrm{aW}/\sqrt{\mathrm{Hz}}$
\\
  $\tau_\mathrm{eff}$ &
  $0.42\pm0.06$~ms &
  $0.40\pm0.06$~ms
\\
  $\eta_\mathrm{opt}$ &
  $0.62\pm0.05$ &
  $0.63\pm0.06$
\\
\hline
\multicolumn{3}{p{.38\textwidth}}{\textit{Note}: Medians and median absolute deviations across TESs on all \nummods{} UHF modules. See text for parameter definitions.}
\end{tabular}
\end{table}

\subsection{Resonators}
Across all \nummods{} modules, an average of 98\% of resonators were identified in vector network analyzer transmission measurements, with a minimum yield of 95\%, corresponding to two modules that each had one defective multiplexer chip replaced with a through chip lacking resonators.
Resonators that are too close in frequency for SMuRF to simultaneously operate (a soft cut on $\Delta f<1.2$~MHz and a hard cut on $\Delta f<0.3$~MHz) bring the average yield down to 90\%, and cuts on SQUID response amplitude and regularity further decrease the average resonator channel yield to 86\%.

\begin{figure}[!h]
\centering
\includegraphics[width=3.3in]{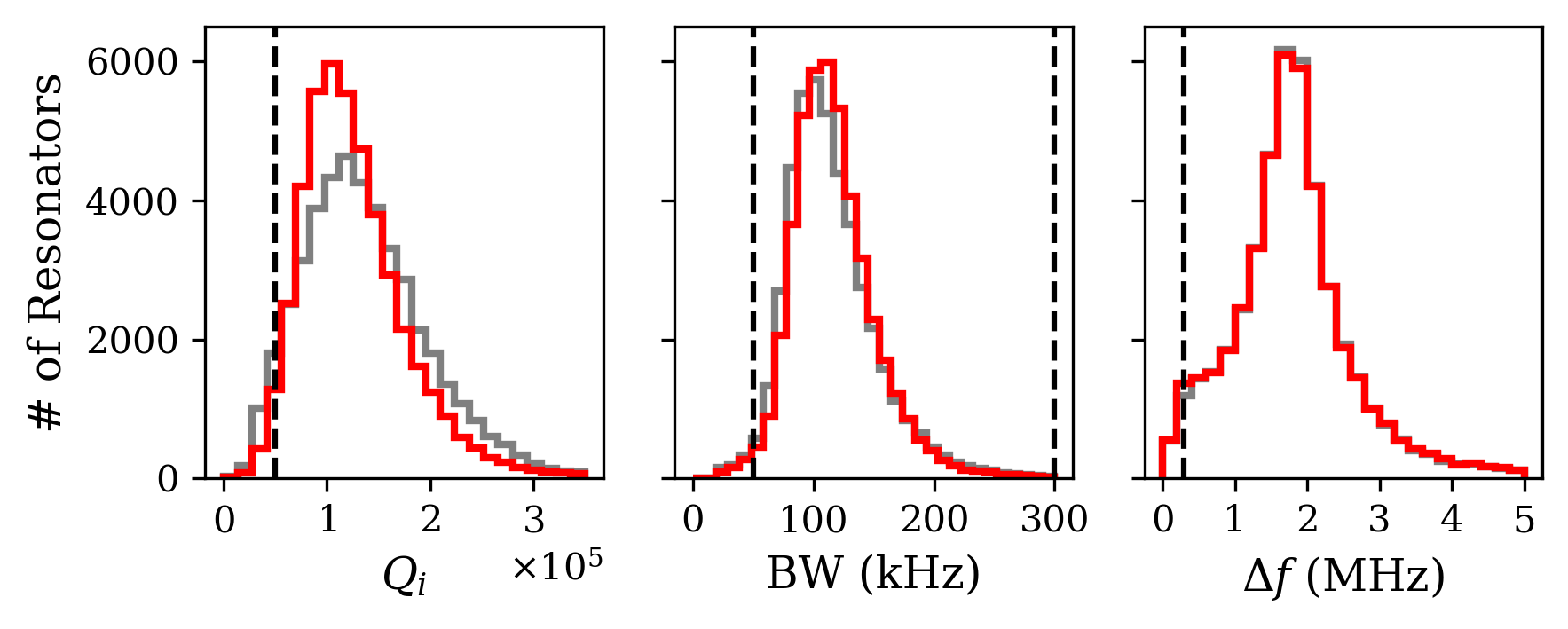}
\caption{
Histograms of resonator internal quality factor $Q_i$, bandwidth, and frequency spacing across all \nummods{} modules before (gray) and after (red) coupling to detectors.
Dashed black lines indicate parameter thresholds consulted during multiplexer chip screening.
When TESs are added to the circuit, the peak of the $Q_i$ distribution shifts down by 9\% and the peak of the bandwidth distribution shifts up by 4\%, while the frequency spacings do not change.
}
\label{fig:resonator_params}
\end{figure}

\begin{figure}[!h]
\centering
\includegraphics[width=3.3in]{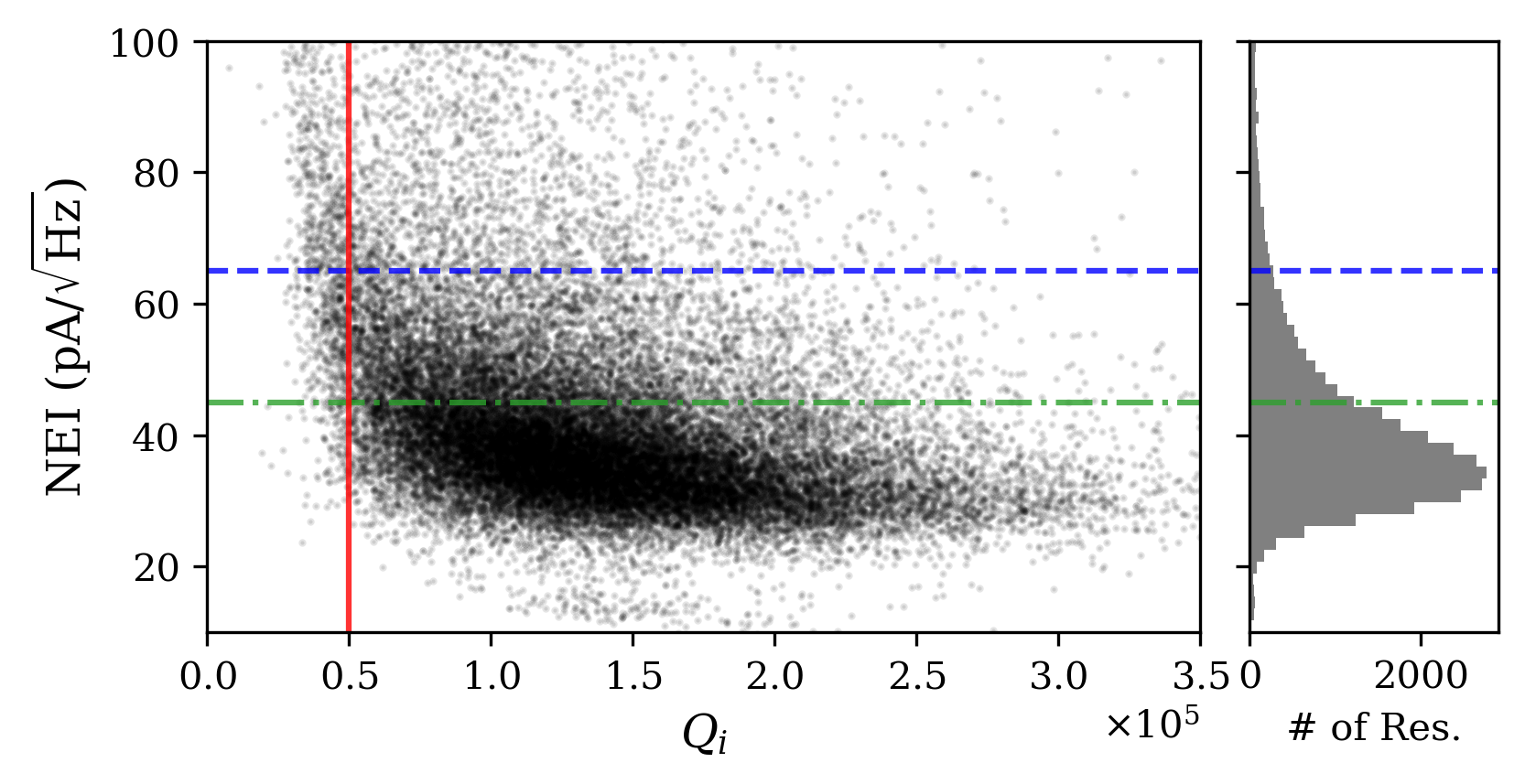}
\caption{
Readout white noise level (NEI) plotted against internal quality factor $Q_i$ for resonator channels across the \nummods{} multiplexing modules.
The blue dashed line indicates the baseline target of 65~pA$/\sqrt{\mathrm{Hz}}$, and the green dot-dash line indicates the goal target of 45~pA$/\sqrt{\mathrm{Hz}}$.
The solid red line marks $Q_i=$50,000, a criterion consulted for accepting or rejecting sets of multiplexer chips.
The distribution of NEI values peaks at 35~pA$/\sqrt{\mathrm{Hz}}$.
}
\label{fig:nei-qi}
\end{figure}

Histograms of measured resonator properties across $\sim$45,000 resonators from all tested modules are shown in Figure~\ref{fig:resonator_params}, both before and after coupling to detectors.
The target frequency spacing for the resonators is 1.8~MHz, which corresponds to the peak of the distribution shown in the rightmost panel of Figure~\ref{fig:resonator_params}, although a significant number of resonators have smaller spacings.
This results in the largest decrease in operable resonator channels discussed above.
The median bandwidth is $\sim$110~kHz, close to the target value of 100~kHz.
The resonator internal quality factor $Q_i$ has the most noticeable change when coupling to detectors, with the distribution narrowing and the median value shifting downward from $1.30\times10^5$ to $1.18\times10^5$.
A degradation in multiplexer $Q_i$ after coupling to detectors has also been noted in a similar design with a higher multiplexing factor of 1820 using a 4--8~GHz readout bandwidth \cite{groh2025}, although in that design the $Q_i$ values nearly halved, causing a substantial increase in overall noise level.
The downward shift in $Q_i$ observed here is only 9\% and has a negligible effect on noise performance.

The measured relation between resonator $Q_i$ and readout noise is plotted in Figure~\ref{fig:nei-qi}.
The criterion $Q_i>$~50,000 was consulted during NIST screening of multiplexer chips \cite{jones2024} when determining to accept or reject whole multiplexer sets, based on the poor noise performance of resonators falling below this when operated with the full readout system.
Ninety-five percent of resonators meet this criterion.
Although there is a tail of high noise channels, the bulk of the readout noise distribution is excellent, peaking at 35~pA$/\sqrt{\mathrm{Hz}}$, with 87\% of channels below the baseline noise target of 65~pA$/\sqrt{\mathrm{Hz}}$ and 68\% of channels below the goal noise target of 45~pA$/\sqrt{\mathrm{Hz}}$.
%These noise targets were defined so as to contribute less than (10\%, 5\%) to the overall detector noise, respectively.
The typical power-to-current responsivity of the UHF TESs is $s_I = 4~\mu \mathrm{V}^{-1}$, meaning these readout noise levels will typically contribute 9~aW$/\sqrt{\mathrm{Hz}}$ to the total detector NEP.

\begin{figure*}[!ht]
\centering
\includegraphics[width=6.4in]{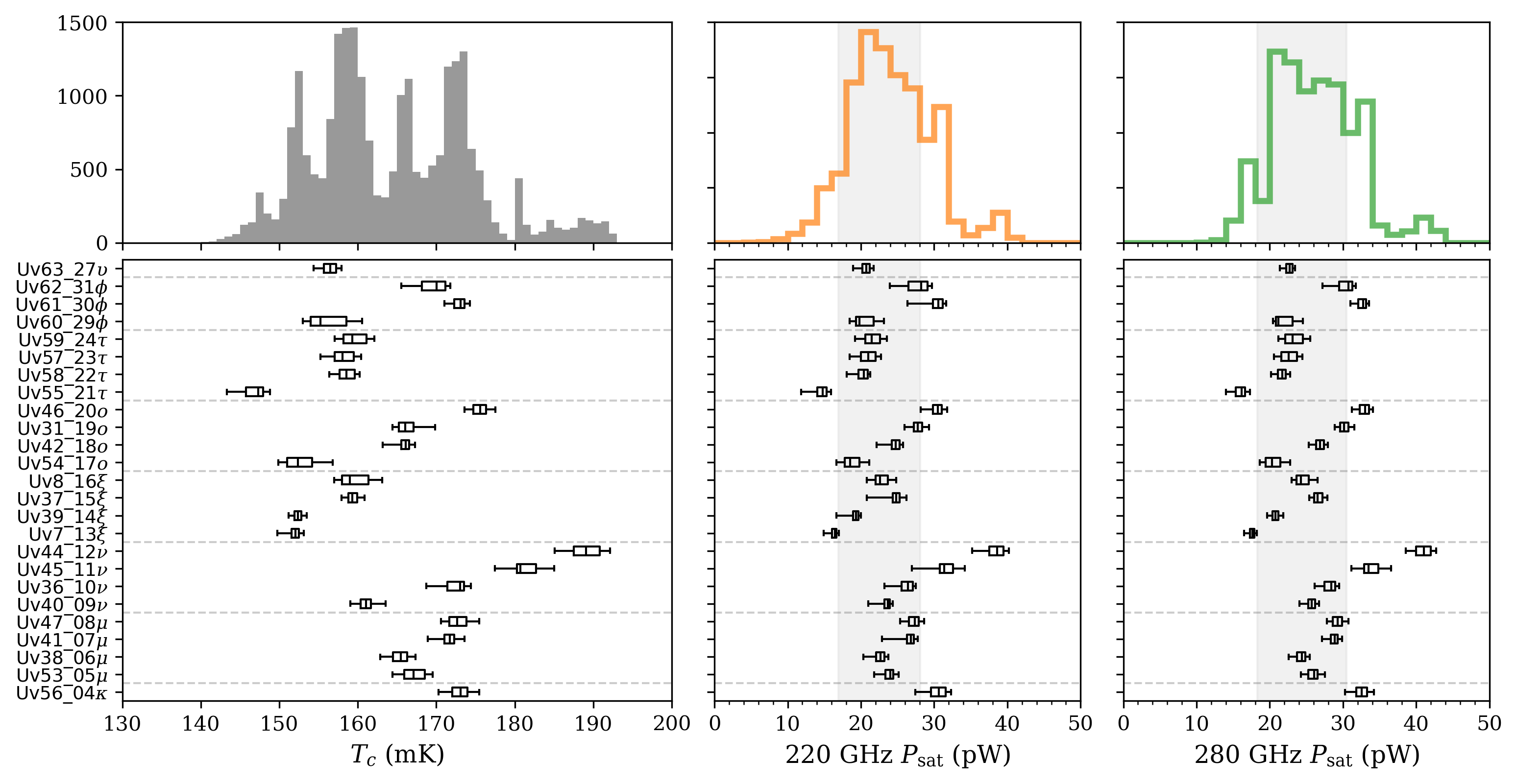}
\caption{
Distributions of TES critical temperature $T_c$ and saturation power $P_\mathrm{sat}$.
The top portion of each plot shows the distribution across all \nummods{} modules.
The lower portions of the plots show the parameter distributions for individual detector modules, where boxes indicate quartiles, and whiskers indicate the central 90\% of the data.
The shaded regions indicate the target $P_\mathrm{sat}$ ranges.
The modules are ordered with the newest detector wafers on top and the oldest at bottom, with dashed lines separating fabrication batches.
}
\label{fig:tc-psat}
\end{figure*}

\subsection{TES Bolometer Properties}
The yielded number of TESs per module is calculated as the number of readout channels exhibiting well-behaved \mbox{$I$-$V$} curves compared to the ideal total of 1756.
Across all UHF modules, the average operable TES yield is 83\%, and the minimum yield is 71\%.
In total, over 36,000 TESs have been measured.
At this stage, all modules meet the baseline yield target of 70\%, and all but six meet the goal yield of 80\%.
While most of the losses originate in the multiplexing readout as discussed above, there are additional losses from inoperable TES bias lines caused by shorts on the DC/RF routing wafer; six detector modules are missing one bias line ($\sim$130--170 TESs) as a result of this.

Figure~\ref{fig:tc-psat} shows the distributions of $T_c$ and $P_\mathrm{sat}$ across all UHF modules, separated by detector wafer fabrication batch.
While the overall median $T_c$ value is close to the target of 160~mK, there is batch-to-batch variation and an unexplained tendency for $T_c$ to rise within a given batch, sometimes significantly, which can also be seen in the data of \cite{dutcher2024}.
As $T_c$ is an important determining factor for the saturation power of the detectors, the impact of this variation in $T_c$ can be seen in the $P_\mathrm{sat}$ distributions in Figure~\ref{fig:tc-psat}.
The saturation powers of the detectors are targeted to be high enough to avoid saturating under predicted typical observing conditions, while not being too high so as to incur an excess noise penalty (see \cite{sonka2025} for details on the specification process for the SO detectors).
The optical loading is predicted as being roughly equal in the \uhfone{}~GHz and \uhftwo{}~GHz bands, and so the target $P_\mathrm{sat}$ ranges are quite similar.
The medians of the $P_\mathrm{sat}$ distributions fall near the middle of the target ranges, at (24, 26)~pW for (\uhfone{}, \uhftwo{})~GHz, respectively.
Ninety-one percent of detectors have $P_\mathrm{sat}$ above the lower edge of the targeted range, with one module, ``Uv55," having sufficiently low $P_\mathrm{sat}$ values so as to be rejected for saturation concerns.
Seventy-seven percent of detectors have $P_\mathrm{sat}$ below the upper edge of the targeted range, although this is a softer bound than the lower edge.
Only module ``Uv44" has a high enough $P_\mathrm{sat}$ to be rejected due to the resulting high detector noise.
Therefore 23 modules have $P_\mathrm{sat}$ values satisfying overall performance requirements, surpassing the 19 UHF modules required to populate the fielded instruments.
%the joint lower and upper is satisfied by 68\% of detectors.

\begin{figure}[!ht]
\centering
\includegraphics[width=3.4in]{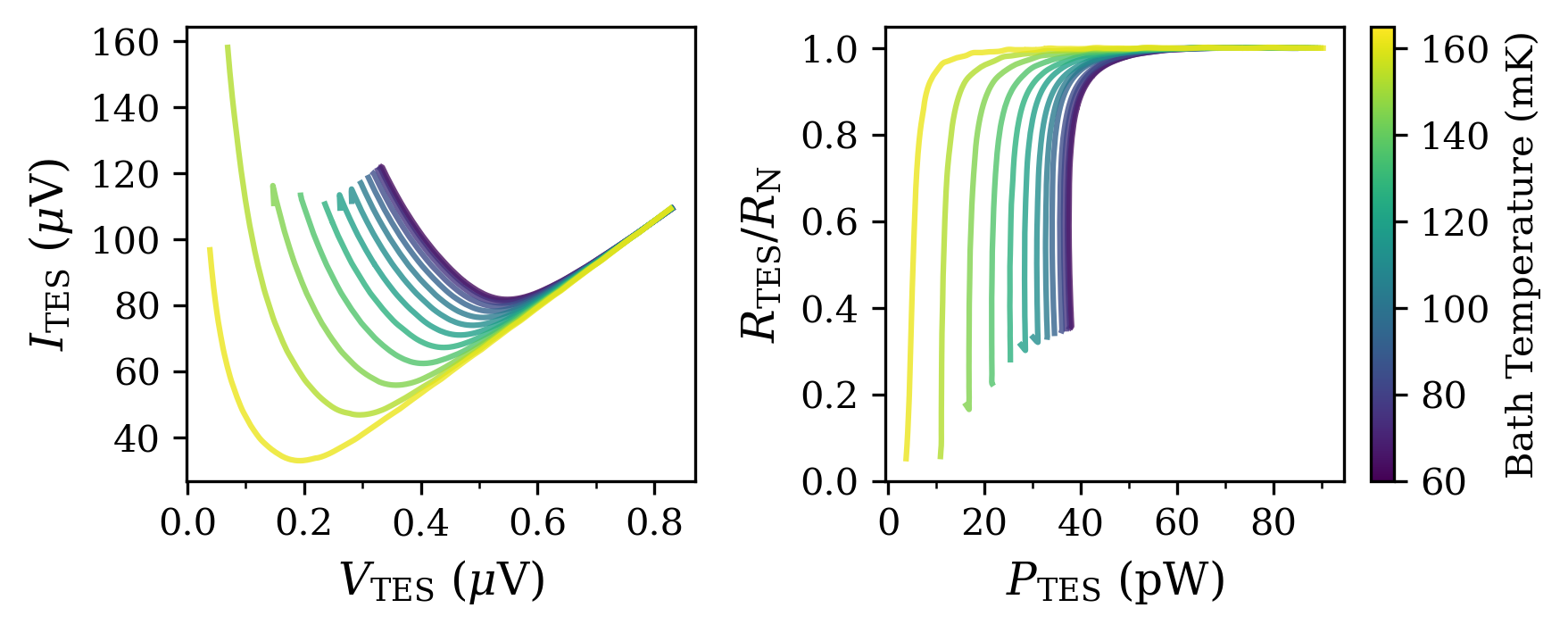}
\caption{
\mbox{$I$-$V$} and corresponding \mbox{$R$-$P$} curves as a function of bath temperature for one TES with $T_c=172$~mK on a high-$P_\mathrm{sat}$ detector module.
The curves extend to the lowest $R_\mathrm{TES}/R_N$ value where the TES can be stably operated.
}
\label{fig:iv}
\end{figure}

The high $P_\mathrm{sat}$ values and fast thermal time  constants (target range $0.2~\mathrm{ms}<\tau_\mathrm{eff}<0.6~\mathrm{ms}$) of the UHF detectors limit the region of the transition where the TESs can be operated stably (see, e.g. \cite{irwin2005} for a discussion).
Figure~\ref{fig:iv} shows a series of \mbox{$I$-$V$} curves from one detector on a high-$P_\mathrm{sat}$ module taken over a range of bath temperatures.
At the coldest bath temperature, when the electrical bias power on the TES ($P_\mathrm{TES}$) is highest, this TES can only be lowered to $0.36R_N$ before the electrothermal feedback loop becomes unstable.
If $P_\mathrm{TES}$ is lowered, either through raising the bath temperature or by external optical loading, the TES can be operated at lower points in the transition.
Empirically, we observe this instability across modules to disappear once $P_\mathrm{TES} \lesssim 15$~pW, which should be the case when operating on-sky for all but the highest-$P_\mathrm{sat}$ modules on the best-weather days.
We therefore do not expect this instability to degrade detector operability.

\begin{figure}[!ht]
\centering
\includegraphics[width=3.3in]{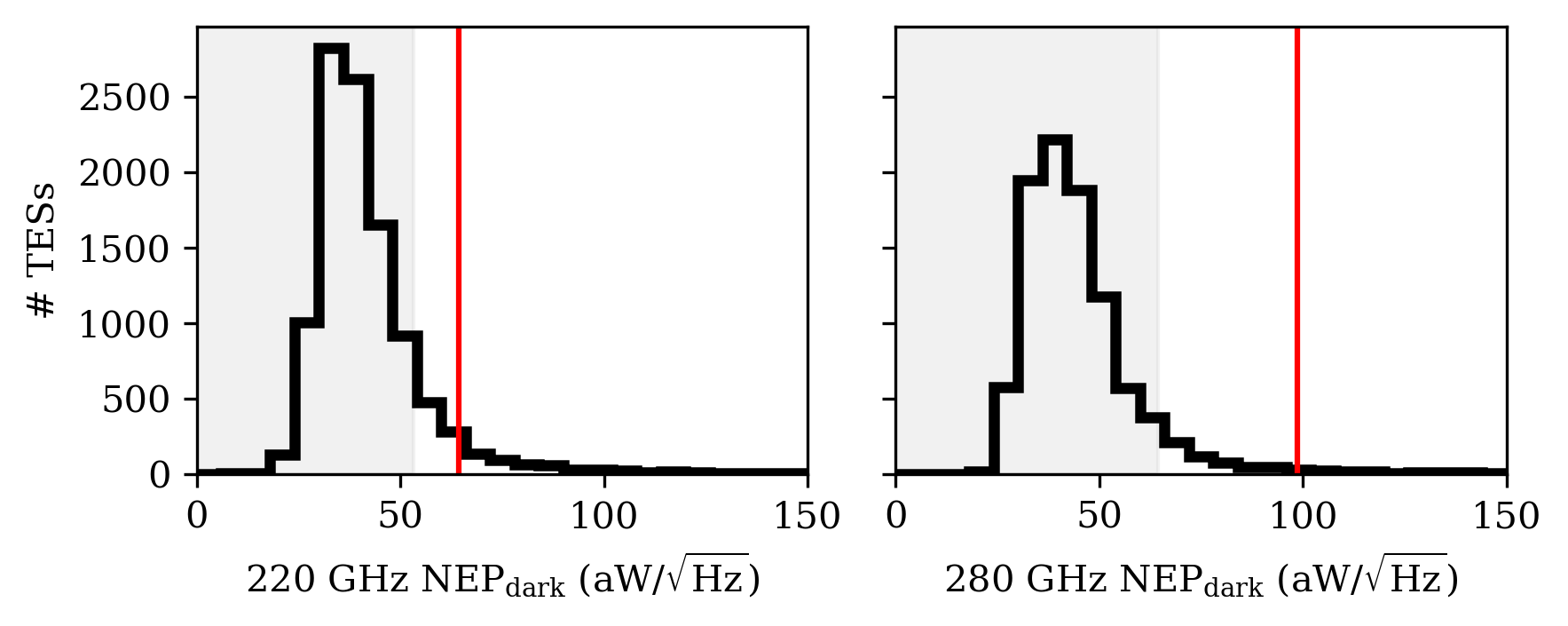}
\caption{
Histograms of detector noise-equivalent power (NEP), measured in-transition at 0.4$R_N$--0.6$R_N$ with no optical loading.
The shaded region extends from zero to the calculated NEP level for detectors meeting baseline parameter targets.
The red lines indicate the predicted photon contributions to NEP for detectors installed on the LAT.
}
\label{fig:nep}
\end{figure}

The noise-equivalent power (NEP) values for detectors measured in the lab are shown in Figure~\ref{fig:nep}, with the data and target ranges both corresponding to masked detectors, i.e. no photon-noise contribution.
The desired upper bounds on NEP are (53, 64)~aW$/\sqrt{\mathrm{Hz}}$ at (\uhfone{}, \uhftwo{})~GHz, while the medians of the measured distributions are (39, 42) aW$/\sqrt{\mathrm{Hz}}$, respectively.
Approximately 89\% of detectors have noise performance better than requirements.
Also plotted in Figure~\ref{fig:nep} are the expected photon noise contributions, which lie above the main distribution of detector noise values, meaning the detectors should achieve background-limited performance on the sky.

\begin{figure}[!ht]
\centering
\includegraphics[width=3.3in]{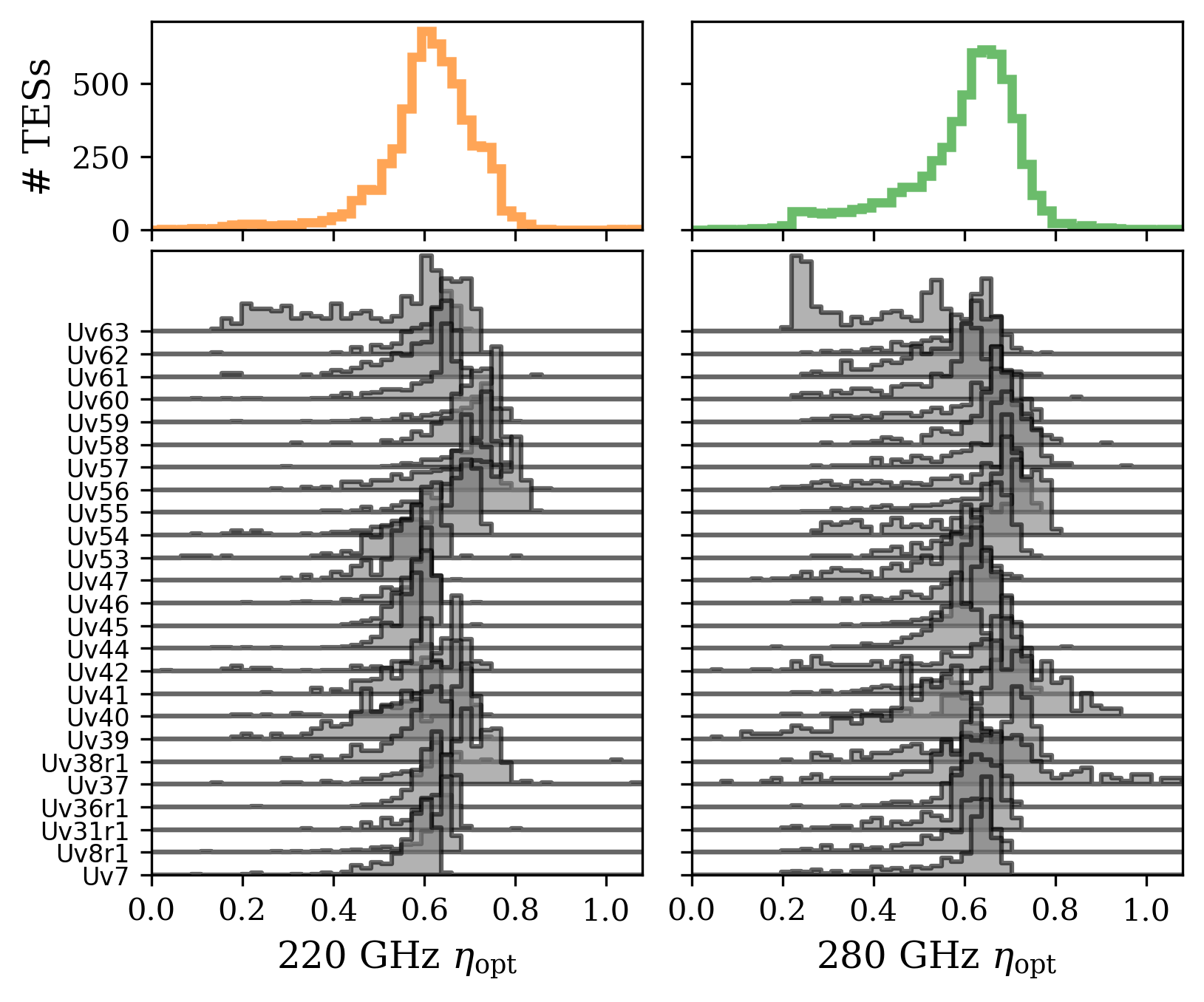}
\caption{
Histograms of optical efficiency values for each module (bottom) and across all modules (top).
Only one-third of each detector module is represented, as the other two-thirds are masked during the measurement.
}
\label{fig:opteff}
\end{figure}

The optical efficiencies of the detectors are plotted in Figure~\ref{fig:opteff}.
The median (interquartile range) for the two observing bands are 0.62 (0.57--0.67) for \uhfone{}~GHz and 0.63 (0.54--0.68) for \uhftwo{}~GHz.
While there is not a strictly defined requirement on detector optical efficiency itself, with the ultimate metric coming from array-averaged noise-equivalent temperature values as measured on the sky, a value of $\eta_\mathrm{opt}\sim0.5$ indicates adequate performance for the UHF detectors.
This metric is surpassed by (89\%, 81\%) of (\uhfone{}, \uhftwo{})~GHz detectors, and overall UHF detector performance should exceed the baseline projections in \cite{thesimonsobservatorycollaboration2019a, thesimonsobservatorycollaboration2025}.

There is a tail of low optical efficiency values similar to that seen in the MF detector wafers \cite{dutcher2024}.
Each UHF detector module has an average of (4\%, 10\%) of (\uhfone{}, \uhftwo{})~GHz detectors with $\eta_\mathrm{opt}<0.4$, although the module ``Uv63" is a clear outlier in this regard, with (35\%, 57\%) of detectors below this threshold, respectively.
These low efficiency tails are known to be exacerbated by post-fabrication thermal re-annealing of the detector wafer, a process used to raise the $T_c$ of the AlMn TESs \cite{li2016} if the initial annealing step \cite{duff2024} did not achieve a high-enough $T_c$, although the exact mechanism by which optical efficiency is impacted by this is still under investigation.
Module ``Uv63" contains the only UHF detector wafer that required this processing step, explaining its outlier status, and as a consequence it will not be deployed to the field.
We will avoid re-annealing on any future detector wafers.

\section{Conclusion}

We have presented the pre-deployment laboratory characterization of \nummods{} UHF detector modules for the Simons Observatory, with over 36,000 TES devices tested.
The modules meet SO performance requirements for yield, optical efficiency, and noise.
The best-performing 19 modules will be installed into SO telescopes in Chile:
thirteen have already been deployed across the LAT and UHF SAT, with another six modules soon to be deployed in an upcoming upgrade to the LAT.
All SO instruments have achieved first-light, and on-sky characterization of the UHF and MF detector performance, including noise-equivalent temperature values, will be addressed in upcoming papers covering telescope commissioning.

\section*{Acknowledgments}
We would like to thank Martina Macakova for her indispensable expertise in wire bonding and mechanical assembly of all SO detector modules, and Kasar Profit and Samuel Morgan for their essential work on detector module installation and cryostat operation.
This work was supported in part by a grant from the Simons Foundation (Award \#457687, B.K.). This work was supported by the U.S. National Science Foundation (Award No. 2153201).

%\begin{thebibliography}{1}
\bibliographystyle{IEEEtran}
\bibliography{LTD2025}

@article{dober2021,
  title = {A Microwave {{SQUID}} Multiplexer Optimized for Bolometric Applications},
  author = {Dober, B. and Ahmed, Z. and Arnold, K. and Becker, D. T. and Bennett, D. A. and Connors, J. A. and Cukierman, A. and D'Ewart, J. M. and Duff, S. M. and Dusatko, J. E. and Frisch, J. C. and Gard, J. D. and Henderson, S. W. and Herbst, R. and Hilton, G. C. and Hubmayr, J. and Li, Y. and Mates, J. A. B. and McCarrick, H. and Reintsema, C. D. and {Silva-Feaver}, M. and Ruckman, L. and Ullom, J. N. and Vale, L. R. and Van Winkle, D. D. and Vasquez, J. and Wang, Y. and Young, E. and Yu, C. and Zheng, K.},
  year = {2021},
  month = feb,
  journal = {Applied Physics Letters},
  volume = {118},
  number = {6},
  pages = {062601},
  doi = {10.1063/5.0033416}
}

@article{duff2024,
  title = {The {{Simons Observatory}}: {{Production-Level Fabrication}} of the {{Mid-}} and {{Ultra-High-Frequency Wafers}}},
  shorttitle = {The {{Simons Observatory}}},
  author = {Duff, Shannon M. and Austermann, Jason and Beall, James A. and Daniel, David P. and Hubmayr, Johannes and Jaehnig, Greg C. and Johnson, Bradley R. and Jones, Dante and Link, Michael J. and Lucas, Tammy J. and Sonka, Rita F. and Staggs, Suzanne T. and Ullom, Joel and Wang, Yuhan},
  year = {2024},
  month = jul,
  journal = {J Low Temp Phys},
  volume = {216},
  number = {1},
  pages = {135--143},
  doi = {10.1007/s10909-024-03117-x}
}

@article{dutcher2024,
  title = {The {{Simons Observatory}}: {{Large-Scale Characterization}} of 90/150 {{GHz TES Detector Modules}}},
  shorttitle = {The {{Simons Observatory}}},
  author = {Dutcher, Daniel and Duff, Shannon M. and Groh, John C. and Healy, Erin and Hubmayr, Johannes and Johnson, Bradley R. and Jones, Dante and Keller, Ben and Lin, Lawrence T. and Link, Michael J. and Lucas, Tammy J. and Morgan, Samuel and Seino, Yudai and Sonka, Rita F. and Staggs, Suzanne T. and Wang, Yuhan and Zheng, Kaiwen},
  year = {2024},
  month = feb,
  journal = {J Low Temp Phys},
  volume = {214},
  number = {3},
  pages = {247--255},
  doi = {10.1007/s10909-023-03045-2}
}

@misc{groh2025,
  title = {Demonstration of a 1820 Channel Multiplexer for Transition-Edge Sensor Bolometers},
  author = {Groh, J. C. and Ahmed, Z. and Austermann, J. and Beall, J. and Daniel, D. and Duff, S. M. and Henderson, S. W. and Hubmayr, J. and Lew, R. and Link, M. and Lucas, T. J. and Mates, J. A. B. and {Silva-Feaver}, M. and Singh, R. and Ullom, J. and Vale, L. and Lanen, J. Van and Vissers, M. and Yu, C.},
  year = {2025},
  month = jul,
  number = {arXiv:2507.10929},
  eprint = {2507.10929},
  primaryclass = {physics},
  publisher = {arXiv},
  doi = {10.48550/arXiv.2507.10929},
  archiveprefix = {arXiv}
}

@article{healy2020,
  title = {Assembly Development for the {{Simons Observatory}} Focal Plane Readout Module},
  author = {Healy, Erin and Ali, Aamir M. and Arnold, Kam and Austermann, Jason E. and Beall, James A. and Bruno, Sarah Marie and Choi, Steve K. and Connors, Jake and Cothard, Nicholas F. and Dober, Bradley and Duff, Shannon M. and Galitzki, Nicholas and Hilton, Gene and Ho, Shuay-Pwu Patty and Hubmayr, Johannes and Johnson, Bradley R. and Li, Yaqiong and Link, Michael J. and Lucas, Tammy J. and McCarrick, Heather and Niemack, Michael D. and {Silva-Feaver}, Maximiliano and Sonka, Rita F. and Staggs, Suzanne and Vavagiakis, Eve M. and Vissers, Michael R. and Wang, Yuhan and Wollack, Edward J. and Xu, Zhilei and Westbrook, Benjamin and Zheng, Kaiwen},
  year = {2020},
  month = dec,
  journal = {Millimeter, Submillimeter, and Far-Infrared Detectors and Instrumentation for Astronomy IX},
  volume = {11453},
  pages = {224--235},
  doi = {10.1117/12.2561743}
}

@article{healy2022,
  title = {The {{Simons Observatory}} 220 and 280~{{GHz Focal-Plane Module}}: {{Design}} and {{Initial Characterization}}},
  shorttitle = {The {{Simons Observatory}} 220 and 280~{{GHz Focal-Plane Module}}},
  author = {Healy, Erin and Dutcher, Daniel and Atkins, Zachary and Austermann, Jason and Choi, Steve K. and Duell, Cody J. and Duff, Shannon and Galitzki, Nicholas and Huber, Zachary B. and Hubmayr, Johannes and Johnson, Bradley R. and McCarrick, Heather and Niemack, Michael D. and Sonka, Rita and Staggs, Suzanne T. and Vavagiakis, Eve and Wang, Yuhan and Xu, Zhilei and Zheng, Kaiwen},
  year = {2022},
  month = dec,
  journal = {J Low Temp Phys},
  volume = {209},
  number = {5},
  pages = {815--823},
  doi = {10.1007/s10909-022-02788-8}
}

@article{henderson2016,
  title = {Advanced {{ACTPol Cryogenic Detector Arrays}} and {{Readout}}},
  author = {Henderson, S. W. and Allison, R. and Austermann, J. and Baildon, T. and Battaglia, N. and Beall, J. A. and Becker, D. and De Bernardis, F. and Bond, J. R. and Calabrese, E. and Choi, S. K. and Coughlin, K. P. and Crowley, K. T. and Datta, R. and Devlin, M. J. and Duff, S. M. and Dunkley, J. and D{\"u}nner, R. and {van Engelen}, A. and Gallardo, P. A. and Grace, E. and Hasselfield, M. and Hills, F. and Hilton, G. C. and Hincks, A. D. and Hlo{\^z}ek, R. and Ho, S. P. and Hubmayr, J. and Huffenberger, K. and Hughes, J. P. and Irwin, K. D. and Koopman, B. J. and Kosowsky, A. B. and Li, D. and McMahon, J. and Munson, C. and Nati, F. and Newburgh, L. and Niemack, M. D. and Niraula, P. and Page, L. A. and Pappas, C. G. and Salatino, M. and Schillaci, A. and Schmitt, B. L. and Sehgal, N. and Sherwin, B. D. and Sievers, J. L. and Simon, S. M. and Spergel, D. N. and Staggs, S. T. and Stevens, J. R. and Thornton, R. and Van Lanen, J. and Vavagiakis, E. M. and Ward, J. T. and Wollack, E. J.},
  year = {2016},
  month = aug,
  journal = {J Low Temp Phys},
  volume = {184},
  number = {3},
  pages = {772--779},
  doi = {10.1007/s10909-016-1575-z}
}

@incollection{irwin2005,
  title = {Transition-{{Edge Sensors}}},
  booktitle = {Cryogenic {{Particle Detection}}},
  author = {Irwin, K.D. and Hilton, G.C.},
  year = {2005},
  pages = {63--150},
  publisher = {Springer Berlin Heidelberg},
  doi = {10.1007/10933596_3},
  isbn = {978-3-540-31478-3}
}

@article{jones2024,
  title = {Qualification of {{Microwave SQUID Multiplexer Chips}} for {{Simons Observatory}}},
  author = {Jones, Dante and Singh, Robinjeet and Austermann, Jason and Beall, J. A. and Daniel, David and Duff, Shannon M. and Dutcher, Daniel and Groh, John and Hubmayr, Johannes and Johnson, Bradley R. and Lew, Richard and Link, Michael J. and Lucas, Tammy J. and Mates, John A. B. and Staggs, Suzanne and Ullom, Joel and Vale, Leila and Van Lanen, Jeffery and Vissers, Michael and Wang, Yuhan},
  year = {2024},
  month = jul,
  journal = {J Low Temp Phys},
  volume = {216},
  number = {1},
  pages = {50--56},
  doi = {10.1007/s10909-024-03102-4}
}

@article{li2016,
  title = {{{AlMn Transition Edge Sensors}} for {{Advanced ACTPol}}},
  author = {Li, Dale and Austermann, Jason E. and Beall, James A. and Becker, Daniel T. and Duff, Shannon M. and Gallardo, Patricio A. and Henderson, Shawn W. and Hilton, Gene C. and Ho, Shuay-Pwu and Hubmayr, Johannes and Koopman, Brian J. and McMahon, Jeffrey J. and Nati, Federico and Niemack, Michael D. and Pappas, Christine G. and Salatino, Maria and Schmitt, Benjamin L. and Simon, Sara M. and Staggs, Suzanne T. and Van Lanen, Jeff and Ward, Jonathan T. and Wollack, Edward J.},
  year = {2016},
  month = jul,
  journal = {J Low Temp Phys},
  volume = {184},
  number = {1},
  pages = {66--73},
  doi = {10.1007/s10909-016-1526-8}
}

@phdthesis{mates2011,
  title = {{The Microwave SQUID Multiplexer}},
  author = {Mates, John Arthur Benson},
  year = {2011},
  school = {University of Colorado}
}

@article{mccarrick2021a,
  title = {The {{Simons Observatory Microwave SQUID Multiplexing Detector Module Design}}},
  author = {McCarrick, Heather and Healy, Erin and Ahmed, Zeeshan and Arnold, Kam and Atkins, Zachary and Austermann, Jason E. and Bhandarkar, Tanay and Beall, James A. and Bruno, Sarah Marie and Choi, Steve K. and Connors, Jake and Cothard, Nicholas F. and Crowley, Kevin D. and Dicker, Simon and Dober, Bradley and Duell, Cody J. and Duff, Shannon M. and Dutcher, Daniel and Frisch, Josef C. and Galitzki, Nicholas and Gralla, Megan B. and Gudmundsson, Jon E. and Henderson, Shawn W. and Hilton, Gene C. and Ho, Shuay-Pwu Patty and Huber, Zachary B. and Hubmayr, Johannes and Iuliano, Jeffrey and Johnson, Bradley R. and Kofman, Anna M. and Kusaka, Akito and Lashner, Jack and Lee, Adrian T. and Li, Yaqiong and Link, Michael J. and Lucas, Tammy J. and Lungu, Marius and Mates, J. A. B. and McMahon, Jeffrey J. and Niemack, Michael D. and {Orlowski-Scherer}, John and Seibert, Joseph and {Silva-Feaver}, Maximiliano and Simon, Sara M. and Staggs, Suzanne and Suzuki, Aritoki and Terasaki, Tomoki and Thornton, Robert and Ullom, Joel N. and Vavagiakis, Eve M. and Vale, Leila R. and Lanen, Jeff Van and Vissers, Michael R. and Wang, Yuhan and Wollack, Edward J. and Xu, Zhilei and Young, Edward and Yu, Cyndia and Zheng, Kaiwen and Zhu, Ningfeng},
  year = {2021},
  month = nov,
  journal = {ApJ},
  volume = {922},
  number = {1},
  pages = {38},
  publisher = {American Astronomical Society},
  doi = {10.3847/1538-4357/ac2232}
}

@phdthesis{sonka2025,
  title = {The {{Simons Observatory Detectors}}: {{From Specification}} to {{Sky}}.},
  shorttitle = {The {{Simons Observatory Detectors}}},
  author = {Sonka, Rita Frances},
  year = {2025},
  school = {Princeton University}
}

@article{thesimonsobservatorycollaboration2019a,
  title = {The {{Simons Observatory}}: Science Goals and Forecasts},
  shorttitle = {The {{Simons Observatory}}},
  author = {{The Simons Observatory Collaboration}},
  year = {2019},
  month = feb,
  journal = {J. Cosmol. Astropart. Phys.},
  volume = {2019},
  number = {02},
  pages = {056},
  doi = {10.1088/1475-7516/2019/02/056}
}

@misc{thesimonsobservatorycollaboration2025,
  title = {The {{Simons Observatory}}: {{Science Goals}} and {{Forecasts}} for the {{Enhanced Large Aperture Telescope}}},
  shorttitle = {The {{Simons Observatory}}},
  author = {{The Simons Observatory Collaboration}},
  year = {2025},
  month = mar,
  number = {arXiv:2503.00636},
  eprint = {2503.00636},
  primaryclass = {astro-ph},
  publisher = {arXiv},
  doi = {10.48550/arXiv.2503.00636},
  archiveprefix = {arXiv}
}

@article{wang2022a,
  title = {Simons {{Observatory Focal-Plane Module}}: {{In-lab Testing}} and {{Characterization Program}}},
  shorttitle = {Simons {{Observatory Focal-Plane Module}}},
  author = {Wang, Yuhan and Zheng, Kaiwen and Atkins, Zachary and Austermann, Jason and Bhandarkar, Tanay and Choi, Steve K. and Duff, Shannon M. and Dutcher, Daniel and Galitzki, Nicholas and Healy, Erin and Huber, Zachary B. and Hubmayr, Johannes and Johnson, Bradley R. and Lashner, Jack and Li, Yaqiong and McCarrick, Heather and Niemack, Michael D. and Seibert, Joseph and {Silva-Feaver}, Maximiliano and Sonka, Rita and Staggs, Suzanne T. and Vavagiakis, Eve and Xu, Zhilei},
  year = {2022},
  month = dec,
  journal = {J Low Temp Phys},
  volume = {209},
  number = {5},
  pages = {944--952},
  doi = {10.1007/s10909-022-02870-1}
}

@article{whipps2023,
  title = {A {{High-Capacity Microwave SQUID Multiplexer Chip Screening System}}},
  author = {Whipps, Zachary and Connors, Jake A. and Dober, Bradley J. and Hubmayr, Johannes and Denison, Edward V. and Vale, Leila R. and Hilton, Gene and Groh, John and Wheeler, Caleb and Gao, Jiansong and Austermann, Jason E. and Mates, J. A. B. and Ullom, Joel N. and Duff, Shannon M. and Johnson, Bradley R. and Wang, Yuhan and Zheng, Kaiwen},
  year = {2023},
  month = jun,
  journal = {J Low Temp Phys},
  volume = {211},
  number = {5},
  pages = {330--337},
  doi = {10.1007/s10909-023-02954-6}
}

@article{yu2023,
  title = {{{SLAC Microresonator RF}} ({{SMuRF}}) {{Electronics}}: {{A}} Tone-Tracking Readout System for Superconducting Microwave Resonator Arrays},
  shorttitle = {{{SLAC Microresonator RF}} ({{SMuRF}}) {{Electronics}}},
  author = {Yu, Cyndia and Ahmed, Zeeshan and Frisch, Josef C. and Henderson, Shawn W. and {Silva-Feaver}, Max and Arnold, Kam and Brown, David and Connors, Jake and Cukierman, Ari J. and D'Ewart, J. Mitch and Dober, Bradley J. and Dusatko, John E. and Haller, Gunther and Herbst, Ryan and Hilton, Gene C. and Hubmayr, Johannes and Irwin, Kent D. and Kuo, Chao-Lin and Mates, John A. B. and Ruckman, Larry and Ullom, Joel and Vale, Leila and Van Winkle, Daniel D. and Vasquez, Jesus and Young, Edward},
  year = {2023},
  month = jan,
  journal = {Review of Scientific Instruments},
  volume = {94},
  number = {1},
  eprint = {2208.10523},
  primaryclass = {astro-ph, physics:physics},
  pages = {014712},
  doi = {10.1063/5.0125084},
  archiveprefix = {arXiv}
}
%\end{thebibliography}

\end{document}